\documentclass[twocolumn,preprintnumbers,amsmath,amssymb,superscriptaddress, altaffilletter, subeqn,prl, maxbibnames=99]{revtex4-1}

\usepackage{physics}
\usepackage{dcolumn}
\usepackage{bm}					
\usepackage{graphicx}
\usepackage{amsmath}

\begin{document}


\title{Tunable magneto-optical properties in MoS$_2$ via defect-induced exciton transitions}


\author{Tomer Amit}

\affiliation{Department of Molecular Chemistry and Materials Science, Weizmann Institute of Science, Rehovot 7610001, Israel}

\author{Daniel \surname{Hernang\'{o}mez-P\'{e}rez}}

\affiliation{Department of Molecular Chemistry and Materials Science, Weizmann Institute of Science, Rehovot 7610001, Israel}

\author{Galit Cohen}

\affiliation{Department of Molecular Chemistry and Materials Science, Weizmann Institute of Science, Rehovot 7610001, Israel}

\author{Diana Y. Qiu}

\affiliation{Department of Mechanical Engineering and Materials Science, Yale University, New Haven, Connecticut 06511, United States}

\author{Sivan Refaely-Abramson}
 
\affiliation{Department of Molecular Chemistry and Materials Science, Weizmann Institute of Science, Rehovot 7610001, Israel}

\email[Corresponding author:]{sivan.refaely-abramson@weizmann.ac.il}




\begin{abstract}

The presence of chalcogen vacancies in monolayer transition metal dichalcogenides (TMDs) leads to excitons with mixed localized-delocalized character and to reduced valley selectivity. Recent experimental advances in defect design in TMDs allow for a close examination of such mixed exciton states as a function of their degree of circular polarization under external magnetic fields, revealing strongly varying defect-induced magnetic properties. A theoretical understanding of these observations  and their physical origins demands a predictive, structure-sensitive theory. 
In this work, we study the effect of chalcogen vacancies on the exciton magnetic properties in monolayer MoS$_2$. Using many-body perturbation theory, we show how the complex excitonic picture associated with the presence of defects---with reduced valley and spin selectivity due to hybridized electron-hole transitions---leads to structurally-controllable exciton magnetic response. We find a variety of g-factors with  changing magnitudes and sign depending on the exciton energy and character. Our findings suggest a pathway to tune the nature of the excitons- and by that their magneto-optical properties-  through defect architecture.
\end{abstract} 
 
\maketitle

Layered transition metal dichalcogenides (TMDs) show unique optical properties, and their associated excited-state phenomena are widely studied for a broad range of applications~\cite{heinz2010atomically, splendiani2010emerging, mak2016photonics}. Their quasi-two-dimensional nature gives rise to strongly bound excitons~\cite{qiu2013optical, chernikov2014exciton, ugeda2014giant, matsuda2015optical}, with structurally-tunable exciton properties~\cite{wang2018colloquium}. In particular, electron-hole transitions at the K and K' valleys in TMDs are valley-selective ~\cite{mak2012control, cao2012valley, xiao2012coupled}, as reflected in their magneto-optic response~\cite{MacNeill2014, stier2016exciton, li2014valley, srivastava2015valley, ashish2021magneto}, making these systems appealing for applications in spintronics and valleytronics~\cite{mak2014valley, Ye2017optical,  10.1002/qute.202000118}. 
The coupling of spin, valley and optical helicity dictates the exciton decay mechanisms ~\cite{wang2014valley, prazdnichnykh2021control, ochoa2013spin, molina2013effect, ginsberg2020}. The involved intervalley decay mechanisms are associated with the underlying exciton exchange interactions~\cite{glazov2014exciton, hao2016direct}, indirect and light-induced occupation of optically dark excitons~\cite{jiang2018microsecond, robert2020, wang2017plane}, and manipulation of the valley selection rules through structural modifications~\cite{rivera2016valley, kim2017observation}.

Of particular interest is the effect of atomic defects on valley and magnetic exciton properties in TMDs~\cite{moody2018microsecond, barthelmi2020atomistic, mitterreiter2021role, guo2020electronic, lin2016defect}. Defect-induced exciton localization leads to efficient quantum emission~\cite{srivastava2015optically,  schuler2019large, lee2018controlling, chakraborty2015voltage} with intriguing implications for quantum information processing~\cite{ye2019spin}. Chalcogen vacancies, common point defects in TMDs~\cite{barja2019identifying},
introduce both occupied and empty localized states~\cite{barja2019identifying,wang2020spin,refaely2018defect,naik2018substrate, gupta2018two}, as well as highly hybridized localized-delocalized exciton transitions~\cite{refaely2018defect,mitterreiter2021role}.
These lead to a decrease in the degree of exciton valley polarization and a significant reduction of the valley-selective optical properties associated with defect transitions, as was shown from theory~\cite{refaely2018defect} and observed in photoluminescence experiments~\cite{klein2017robust, mitterreiter2021role}.
This change in the valley degree of freedom can be directly detected through polarization-resolved magneto-optical spectroscopy under external magnetic fields. While the magnitude of exciton g-factors in pristine TMDs are found to be around 4~\cite{deilmann2020ab, stier2016exciton, ashish2021magneto}, measurements of defect-associated g-factors give a variety of results, from vanishing magnitudes of zero to greatly enhanced ones on the order of 10~\cite{dang2020identifying, he2015single,wang2020spin}. The effect of defect states on the magneto-optic phenomena is thus complex and demands careful analysis.

A comprehensive and predictive theoretical understanding of defect-induced optical excitations and their magnetic properties is therefore essential. Recent advances allow for first-principles calculations of the orbital magnetic moment~\cite{wozniak2020exciton} and the associated exciton g-factors in pristine TMDs~\cite{deilmann2020ab, forste2020exciton, xuan2020valley}. Previous calculations have considered the effect of defects on magneto-optical properties in TMDs from the perspective of single-particle band transitions~\cite{wang2020spin} or by using a tight-binding BSE approach that does not include the entire Hilbert space of electron-hole transitions~\cite{linhart2019localized}. However, a fully \textit{ab initio} picture of the effect of defect excitons on the magneto-optical properties has  yet to be considered.
Such an understanding can offer a tractable pathway to both identify atomic defects from their magnetic response and use their magneto-optical properties to tune TMD valley selectivity.

In this work, we present a first-principles study of the effect of atomic defects on the magneto-optical properties in monolayer TMDs. We use many-body perturbation theory within the GW-BSE approximation to compute exciton transitions in MoS$_2$ with chalcogen vacancies, and derive the associated magnetic moments and exciton g-factors. Our calculations show diverse exciton magnetic properties stemming from a defect-induced mixing of the electron-hole transitions. These lead to a spectrum of different g-factors associated with defect excitons, ranging from positive to negative values and with vanishing to enhanced magnitudes. We analyze how hybridization between defect and pristine-like states breaks both valley and spin selectivity, allowing for direct optical excitation of states which are optically dark in the pristine case. Our findings reveal how the presence of defects alters the conventional picture of exciton g-factors in TMDs, suggesting tunable, structure-induced magnetic moments through defect design.

\begin{figure}
    \includegraphics[width=1.0\linewidth]{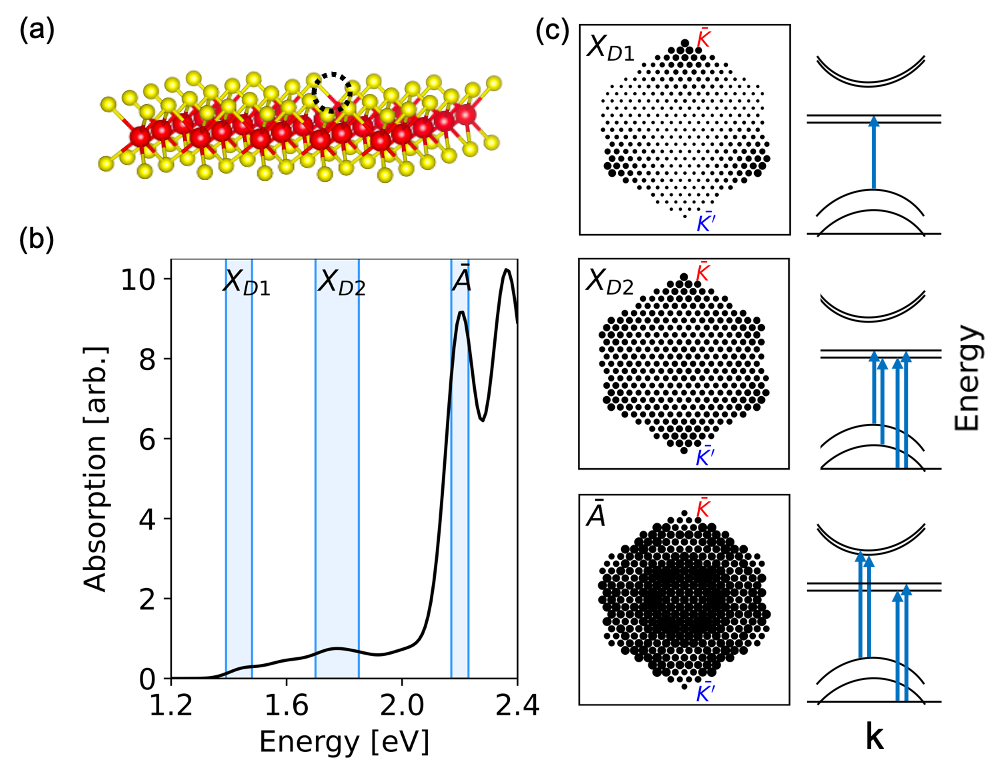}      
    \caption{(a) The studied system of monolayer MoS$_2$ with a single sulphur vacancy in a 5$\times$5 periodic supercell. (b) Calculated GW-BSE absorption spectrum. The energy regions associated with the three broad exciton features analyzed in this work are highlighted in blue: the A-like peak, $\bar{A}$, and two main low-energy defect-induced peaks, $X_{D1}$ and $X_{D2}$. (c) The k-resolved exciton contributions as defined in Eq.~\ref{Af}, summed over the energy windows marked in (b). A schematic representation of the dominant electron-hole transitions is shown for each peak.}\label{fig:13} 
\end{figure}

Figure~\ref{fig:13} shows the studied system, a MoS$_2$ monolayer with chalcogen vacancies, as well as its calculated absorption spectrum and the extent of the dominant exciton peaks in reciprocal space. We use a 5$\times$5 periodic supercell leading to 2\% defect density, as illustrated in Fig.~\ref{fig:13}a (see SI for full computational details). As previously found, this supercell size is required to minimize interactions between vacancies in neighboring periodic images~\cite{refaely2018defect}. We compute the quasiparticle bandstructure using the one-shot G$_0$W$_0$ approach~\cite{hybertsen1986electron,deslippe2012berkeleygw} on top of the DFT (PBE)~\cite{perdew1996generalized} ground state, in a fully-relativistic formalism that explicitly includes both the spatial and spinor degrees of freedom. Exciton states are then computed using the GW plus Bethe Salpeter equation (BSE) approach~\cite{rohlfing1998electron, deslippe2012berkeleygw}, via the solution of the BSE in the Tamm-Dancoff approximation~\cite{rohlfing2000electron}:
\begin{equation}\label{BSE}
\begin{split}
    \left(E_{c\mathbf {k}}-E_{v\mathbf{k}}\right)A^S_{vc\mathbf{k}}+\sum_{v'c'\mathbf{k'}}{\mel{v\mathbf{k};c\mathbf{k}}{K^{eh}}{v'\mathbf{k'};c'\mathbf{k'}}}A^S_{v'c'\mathbf{k'}}
    \\= \Omega^SA^S_{vc\mathbf{k}}.
\end{split}
\end{equation}
Here, $E_{c\mathbf{k}}$ and $E_{v\mathbf{k}}$ are the GW quasiparticle energies of the conduction ($c$) and valence ($v$) bands, with crystal momentum $\mathbf{k}$, while S indexes the exciton state. ${K^{eh}}$ is the BSE electron-hole interaction kernel, $\Omega^S$ is the exciton excitation energy, and $A^S_{vc\mathbf{k}}$ is the electron-hole amplitude of an exciton state,
\begin{equation} \label{exciton_basis}
    S=\sum_{v c\mathbf k}{A^S_{vc\mathbf{k}}\psi_{c\mathbf k}\psi^*_{v\mathbf k}},
\end{equation}
where $\psi_{c\mathbf k}$ ($\psi^*_{v\mathbf k}$) are the electron (hole) wavefunctions. 

The computed GW-BSE absorption spectrum is shown in Fig.~\ref{fig:13}b for right-handed circularly polarized light ($\sigma^+$). Fig.~\ref{fig:13}c shows the momentum-resolved exciton distributions. The labels $\mathrm{\bar{K}}$, $\mathrm{\bar{K}'}$ represent the $K$-valleys in the supercell Brillouin Zone (BZ), which also corresponds to the $K$-valleys of the unit cell. The exciton distributions are weighted by the absorption oscillator strength, $f^{S'}_{\sigma^+}:$
\begin{equation}
     \left|A_{\mathbf{k}}\right|^2=\sum_{vcS'}f^{S'}_{\sigma^+}\left|A_{vc\mathbf{k}}^{S'}\right|^2, \label{Af}
\end{equation} where $S'$ denotes an exciton in a specific energy range associated with three dominant low-lying exciton features (highlighted in blue in Fig.~\ref{fig:13}b): the $X_{D1}$ peak region is primarily composed of optical transitions between a pristine-like valence band to in-gap defect bands at $\bar{K}$, as shown schematically in Fig.~\ref{fig:13}c and discussed in detail below.  At the $X_{D2}$ peak energy, additional exciton contributions appear throughout the BZ, originating from transitions between an occupied defect band at the valence region and unoccupied in-gap defect bands. The defect character leads to reduced magnetic moments compared to the pristine excitons and mixing of transitions between both spin up and down states. The third peak, $\bar{A}$, superficially resembles the pristine A exciton peak, but in the presence of defects, this region includes a mixture of electron-hole transitions that are no allowed in the pristine case. The main contributions arise from transitions from the pristine-like valence band to the pristine-like conduction bands at $\bar{K}$; however, the presence of defects changes the exciton character compared to the pristine case and allows a small contribution from intravalley opposite-spin excitations at $\mathrm{\bar{K}}$, $\mathrm{\bar{K}'}$. Defect-defect transitions also contribute to excitons in this region as well. The resulting mixed nature of the excitons, and the breaking of valley and spin selection rules associated with it, lead to varying exciton g-factors, as we discuss in detail below.

\begin{figure*}
    \includegraphics[width=1.0\linewidth]{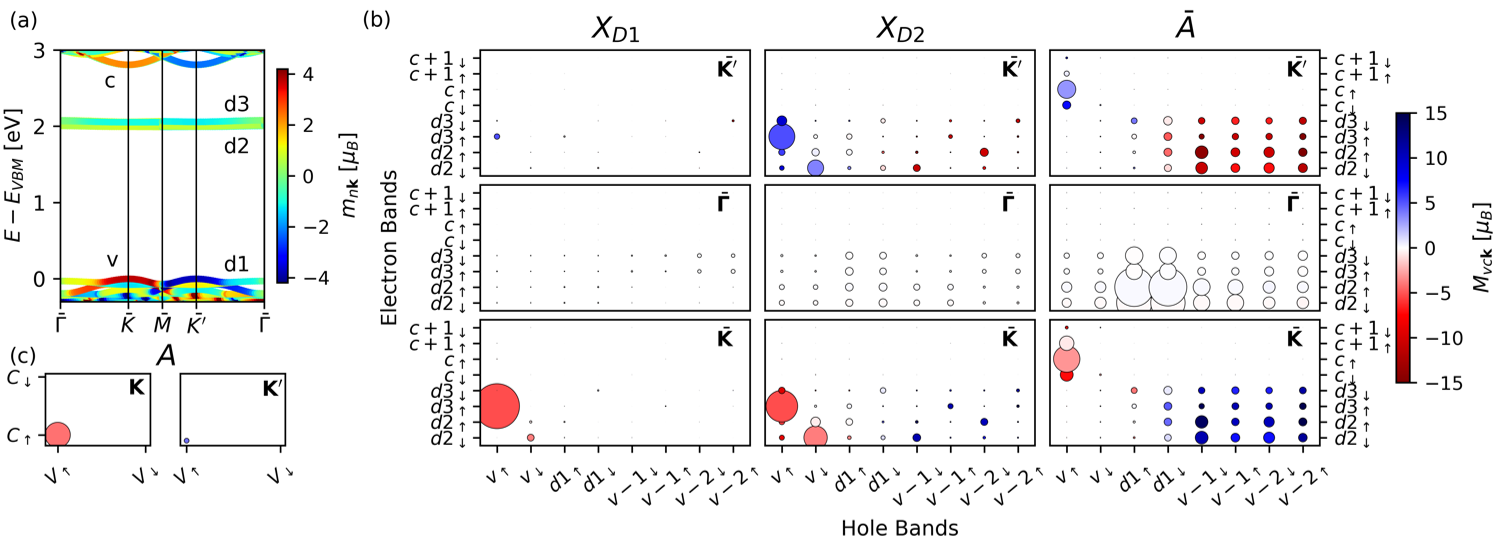}      
    \caption{(a) Calculated quasiparticle bandstructure of the examined MoS$_2$ monolayer with sulphur vacancies. Colors represent the band net magnetic moment, $m_{n\mathbf{k}}$, in units of Bohr's magneton. (b) Analysis of the electron-hole transitions composing the excitons, upon light polarized at the $\sigma^+$ direction at the representative k-points $\mathrm{\bar{K}}$ , $\mathrm{\bar{K'}}$, and $\Gamma$. Dot size represents the relative contribution of each transition, weighted by the oscillator strength of each exciton and summed across the peak energy region ($f^{S'}_{\sigma^+}|A_{vc\mathbf{k}}^{S'}|^2$), for the defect-containing MoS$_2$. Dot colors represents the two-particle magnetic-moment differences, $M_{vc\mathbf{k}}=2\left(m_{c\mathbf{k}}-m_{v\mathbf{k}}\right)$, for each transition. Arrows denote the out-of-plane spin direction at $\bar{K}$. (c) Same analysis as in (b) but for the pristine MoS$_2$ monolayer.}\label{fig:11} 
\end{figure*}

Figure~\ref{fig:11} shows the computed single-particle magnetic moments $m_{n\mathbf{k}}$ (for an electron at state $n$ and k-point $\mathbf{k}$) and the two-particle magnetic-moment differences $M_{vc\mathbf{k}} = 2\left(m_{c\mathbf{k}}-m_{v\mathbf{k}}\right)$ associated with the electron-hole transitions composing the GW-BSE excitons. 
The quasiparticle bandstructure of monolayer MoS$_2$ with chalcogen vacancies, Fig.~\ref{fig:11}a, shows three spin-degenerate localized bands at the gap region: an occupied state ($d1$) within the valence region, which is strongly hybridized with the pristine-like bands at the valence region; and two unoccupied in-gap states ($d2$; $d3$) below the pristine-like conduction band. The spatial localization of these defect bands leads to reduced k-sensitivity and net magnetic moments, with a magnitude of up to $\abs{m_{n\mathbf{k}}}=1~\mu_B$ (see SI for further details).  
The pristine-like valence ($v_\uparrow$) and conduction ($c_\uparrow$) bands around $\mathrm{\bar{K}}$ and $\mathrm{\bar{K'}}$ maintain magnetic moments similar to pristine MoS$_2$ with magnitudes of $3.8 \mu_B$ and $2.2 \mu_B$, respectively (where the sign depends on the valley). Notably, the pristine-like spin-split valence band ($v_\downarrow$) have a reduced magnetic moment due to hybridization with defect bands at the valence region. 

The presence of both defect-localized and pristine-like states around the gap leads to a variety of electron-hole transitions contributing to the excitonic landscape.
Figure~\ref{fig:11}b shows the transitions composing the computed GW-BSE excitons in each of the three energy peak regions defined above, at the representative high-symmetry k-points $\mathrm{\bar{K}}$, $\mathrm{\bar{K'}}$, and $\mathrm{\bar{\Gamma}}$. Each dot represents a transition from a valence band (x-axis) to a conduction band (y-axis). The dot size represents the relative contribution of each transition to the excitons composing the peak, and is weighted by the oscillator strength for optical transitions with right-handed circularly polarized light ($\sigma^+$), namely the summed elements in Eq.~\ref{Af}. The dot color represents the two-particle magnetic-moment difference $M_{vc\mathbf{k}}$, for each electron-hole transition. The factor of 2 stems from the definition of the exciton g-factor in the independent particle approximation at the right-handed ($\sigma^+$) circularly polarized light compared to that with left-handed ($\sigma^-$) circularly polarized light, which we follow in our calculations. 

For comparison, Fig.~\ref{fig:11}c shows the transitions corresponding to the A exciton peak of the pristine system.
We note that the presence of defects changes the identification of spin-allowed and spin-forbidden states. $S_z$ is a good quantum number for pristine MoS$_2$ at the K and K' points, so that for the case of in-plane light polarization, the bright A exciton is only composed of the same-spin $\mathrm{V}_\uparrow \rightarrow \mathrm{C}_\uparrow$ transition, and the opposite-spin $\mathrm{V}_\uparrow \rightarrow \mathrm{C}_\downarrow$ transition is dark (where $\mathrm{V}$, $\mathrm{C}$ denote the pristine valence and conduction bands). 
With chalcogen vacancies included, the defect states no longer have a well defined $S_z$: namely, the spin is no longer a well-defined quantum number even at $\mathrm{\bar{K}}$, $\mathrm{\bar{K'}}$. Hence, while in the pristine system, transitions between up and down spin states are not allowed, in the defect system, they become allowed and are part of the transitions building the optically-bright excitons.

The main electron-hole transition contributing to the lowest exciton peak, $X_{D1}$,  is between the highest pristine-like valence band and a defect in-gap band, $\mathrm{v}_\uparrow\rightarrow\mathrm{d_3}$, at the $\mathrm{\bar{K}}$ point. The two-particle magnetic-moment difference associated with this transition is $M=-5.2 \mu_B$. Two additional small contributions appear in this peak: one for the opposite valley, $\mathrm{\bar{K'}}$, pointing to a slight breaking of valley selectivity in this state; and another for the transition from the opposite-spin valence to an in-gap defect band, $\mathrm{v}_\downarrow\rightarrow\mathrm{d_2}$. In contrast to the $X_{D1}$ peak, the exciton peak $X_{D2}$ has a significantly hybridized nature. On top of the above pristine-like to defect transitions, this energy region includes additional contributions from defect-defect transitions across the BZ.  Despite the fact that the exciton contributions are weighted by the oscillator strength under right circularly polarized light, defect-defect transitions contribute at both valleys, due to the reduced valley selectivity of these localized bands. As a consequence of exciton hybridization, in this energy region we also observe breaking of the valley selectivity for the pristine-like to defect transitions, with opposite signs of the two-particle magnetic-moment difference, $M_{vc\mathbf{k}}$ at $\mathrm{\bar{K}}$ and $\mathrm{\bar{K'}}$. 

The $\bar{A}$ peak region is also strongly hybridized, with large contributions coming from the $\mathrm{\bar{\Gamma}}$ point and its vicinity. This energy range also includes transitions from  pristine-like valence states to defect in-gap states, leading to large two-particle magnetic-moment differences of up to $\abs{15}~\mu_B$. Additional contributions arise from transitions between pristine-like valence and conduction bands at $\mathrm{\bar{K}}$ and $\mathrm{\bar{K'}}$, namely the equivalent transitions to the $A$ exciton in the pristine system. However, due to the above-discussed defect-induced structural changes, transition to both spin-split pristine-like conduction bands are optically allowed within the bright exciton peak. In other words, the optically dark A-peak transition in the pristine monolayer, $\mathrm{V}_\uparrow \rightarrow \mathrm{C}_\downarrow$ (Fig.~\ref{fig:11}c), becomes optically-allowed upon defect presence. Notably, owing to the different spin contributions, while the $\mathrm{v}_\uparrow \rightarrow \mathrm{c}_\downarrow$ transition has $M_{vc\mathbf{k}}=\pm7.3 \mu_B$, for the $\mathrm{v}_\uparrow \rightarrow \mathrm{c}_\uparrow$ transition $M_{vc\mathbf{k}}= \pm3.1 \mu_B$.
Here again we observe the breaking of valley selectivity, allowing these transitions to appear in both $\mathrm{\bar{K}}$ and $\mathrm{\bar{K'}}$ despite the circularly polarized light.

\begin{figure}
    \includegraphics[width=1.0\linewidth]{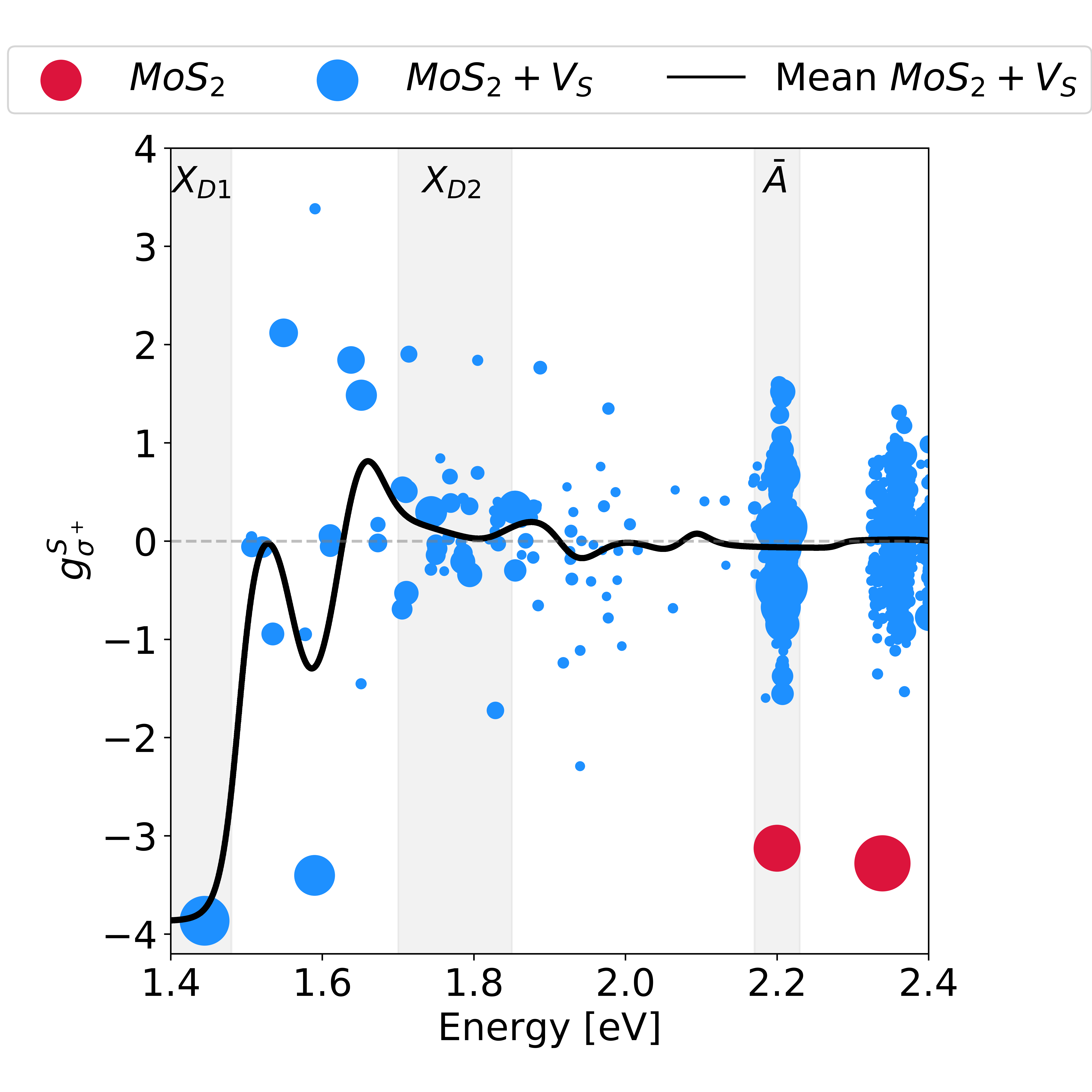}      
    \caption{Calculated exciton g-factors as a function of exciton energy. Dot size is proportional to the computed oscillator strength at the ${\sigma^+}$ polarization direction, $f^{S}_{\sigma^+}$. Red dots show the resulting g-factors for the pristine MoS$_2$. Pristine excitation energies are shifted up by 0.22 eV so that the pristine $A$ peak coincides with the defect-induced $\bar{A}$ peak. Blue dots show the exciton g-factors in MoS$_2$ with sulphur vacancies. The black line shows the mean g-factor, averaging over the calculated exciton g-factors in the defect-containing system. }\label{fig:gfactors9} 
\end{figure}

Finally, we compute how the defect-induced excitonic band mixing affects exciton g-factors in the studied system. We follow a method recently suggested by Deilmann et al.~\cite{deilmann2020ab} for the pristine case, to calculate the exciton absorption g-factor. The two-particle magnetic-moment difference is weighted through the computed BSE exciton coefficients, via: 
\begin{equation} \label{exciton_g_factor}
    g^S = \frac{1}{\mu_B}\sum_{v c\mathbf k}{\left|A^S_{vc\mathbf{k}}\right|^2M_{vc\mathbf k}},
\end{equation}
where $g^S$ is the g-factor of exciton $S$. We then look only at excitons which are bright upon absorption with light polarization along the ${\sigma^+}$ direction, $g^S_{\sigma^+}$.
Figure~\ref{fig:gfactors9} shows the computed exciton g-factors as a function of excitation energy. Blue dots correspond to the GW-BSE exciton g-factors of the defect system, for each exciton state $S$. The size of each dot represents the oscillator strength of the exciton upon $g^S_{\sigma^+}$ light polarization. For comparison, red dots show the computed g-factors for pristine MoS$_2$, of $g^S_{\sigma^+}=-3.1$ and $g^S_{\sigma^+}=-3.3$ for the A and B peaks, respectively, in good correspondence with previous findings~\cite{deilmann2020ab}. Pristine excitation energies are shifted up by 0.22 eV so that the pristine $A$ peak coincides with the defect-induced $\bar{A}$ peak. 

Once defects are introduced, our results show a large variety of exciton g-factors, originating from the mixed transitions discussed above. The lowest exciton state, $X_{D1}$, has a well-defined g-factor of $g^S_{\sigma^+}=-3.9$ due to the relatively selective electron-hole transitions involved. However, for higher excitation energies,  the computed g-factors are of varying magnitudes and with both positive and negative values, depending on the electron-hole transitions constructing them, owing to the mixed valley and spin components shown in Fig.~\ref{fig:11}b. At the $X_{D2}$ peak region, the combination of positive and negative two-particle magnetic-moment differences leads to vanishing g-factors. At the $\bar{A}$ peak energy region, the exciton g-factors range from $g^S_{\sigma^+}=0$ to $\pm 1.6$, with both positive and negative signs present and with a much reduced magnitude compared to the pristine case, due to the defect-induced exciton hybridization. 
The black line shows the mean g-factor along the examined excitation energies, weighted by the exciton brightness, which is equivalent to weighting by the population of excitons excited by right circularly polarized light. Notably, the computed g-factors at the $\bar{A}$ region average to zero due to cancellation of g-factors of opposite signs. This is a direct manifestation of the large reduction of valley selectivity upon the presence of defects.

We note that our results are consistent with previous calculations, which predict only an enhancement of the exciton g-factor~\cite{wang2020spin,linhart2019localized}: if one merely accounts for the electron-hole transitions between the pristine-like valence and the in-gap defect conduction bands at $\mathrm{\bar{K}}$, enhanced g-factor magnitudes of $5-10$ are found. However, we emphasize that this enhancement is only true for a simplified effective-mass picture, which does not include the full spectrum of defect excitations and hybridization between the pristine-like states and the defect states. Once the full spectrum of excitations is taken into account, the computed g-factors are largely spread and generally reduced. It is worth mentioning, however, that our calculations for absorption exciton properties are expected to change for exciton observed in photoluminescence, in which the exciton mixing is expected to reduce within their decay processes. 

In practice, our results imply that in absorbance spectroscopy of the examined system, a Zeeman shift of the absorption resonance will average out. This prediction is consistent with recent experimental findings~\cite{klein2017robust}, in which a drastic reduction of A-exciton valley polarization was observed upon the formation of chalcogen vacancies. This observation directly points to possible tuning of the exciton magnetic response through external effects such as strain, charge, and electric fields, as well as atomic substitutions, all modify the electronic bandstructure and hence the level of exciton state mixing~\cite{linhart2019localized, naik2018substrate, gupta2018two, klein2022electrical, klein2021controlling, hotger2021gate}. For example, electron charging can move the unoccupied defect state position relative to the gap (see SI). Such selective band shifting changes the mixed transitions composing each exciton, and in particular will result in well-isolated defect-defect low-lying excitons with vanishing g-factors, following our analysis above. As such, defect charging can serve as a tuning knob for the magnetic nature of low-lying excitons in the system. However, following our observations above, the defect-defect transitions in this case are still expected to mix with higher transitions, and thus conserve a hybridized nature of the $\bar{A}$ peak. Our analysis of the exciton g-factors further demonstrates how defect-induced changes in the TMD states and the corresponding selection rules lead to optically-allowed excitations that are considered to be spin-forbidden in the pristine case. Such excitonic mixing supports a recently-suggested mechanism of g-factor reduction owing to excitation of spin-forbidden excitons in MoS$_2$~\cite{robert2020}. 


In conclusion, we show that excitonic hybridization between pristine-like and defect states modifies the exciton magnetic properties, leading to a large variety of possible exciton g-factors, with negative, positive or vanishing g-factors as a function of the exciton state and the electron-hole transitions composing it. We further demonstrate that the g-factors at the A-like excitation region average to zero due to these mixed transitions, manifesting the breaking of valley-selectivity and the magnetic response associated with it. Our results thus suggest defect design as a plausible route for tunable magnetic properties in TMDs. 

\textbf{Acknowledgments:} We thank Paulo E. Faria Junior, Alexander W. Holleitner, Christoph Kastl, Andreas V. Stier, Alexander H{\"o}tger, Manish Jain, and Sudipta Kundu for helpful discussions. 
T.A. is supported by the David Lopatie Fellows Program. The work of D.Y.Q was supported by the U.S. Department of Energy, Office of Science, Office of Basic Energy Sciences under Award Number DE-SC0021965. S.R.A. is an incumbent of the Leah Omenn Career Development Chair and acknowledges a Peter and Patricia Gruber Award and an Alon Fellowship.
Computational resources were provided by the Oak Ridge Leadership Computing Facility through the Innovative and Novel Computational Impact on Theory and Experiment (INCITE) program, which is a DOE Office of Science User Facility supported under Contract No. DE-AC05-00OR22725. Additional computational resources were provided by the ChemFarm local cluster at the Weizmann Institute of Science.


%

\end{document}